\input phyzzx
\Frontpage
\centerline{AN INTERACTING GEOMETRY MODEL AND INDUCED GRAVITY}
\vfill
\centerline{Pawel O. Mazur}
\vskip .2in
\centerline{Institute For Fundamental Theory, Physics Department}
\centerline{University of Florida, Gainesville, FL 32611}
\vskip .2in
\centerline{V. P. Nair}
\vskip .2in
\centerline{Columbia University, Physics Department,}
\centerline{New York, N. Y. 10027}
\vskip .5in
\centerline{Gravity Research Foundation Essay, March 1988}
\vskip .1cm
\centerline{Published in GRG 21, 651-658 (1989)}
\vfill
\centerline{Abstract}
\vskip .2in
\par
We propose the theory of quantum gravity with interactions introduced by
topological principle. The fundamental property of such a theory is that its
energy-momentum tensor is an BRST anticommutator. Physical states are elements
of BRST cohomology group. The model with only topological excitations,
introduced recently by Witten is discussed from the point of view of induced
gravity program. We find that the mass scale is induced dynamically
by gravitational instantons. The low energy effective theory has gravitons,
which occur as the collective excitations of geometry, when the metric becomes
dynamical. Applications of cobordism theory to QG are discussed.
\vfill
\endpage
\doublespace
\par
The idea that topology of spacetime must play the fundamental role in any
reasonable theory of Quantum Gravity (QG) was first put forward
by Wheeler$^{1}$. This highly original physical proposal was subsequently taken
seriously by a number of people $^{2,3,4}$, especially after the period
of intensive studies of nonperturbative properties of gauge theories
(instantons, monopoles). The concept of gravitational instanton was
introduced by Hawking$^4$. The classical role of instantons is to break 
classical symmetries of the action functional for a given physical system.
A beautiful example of this phenomenon is the t`Hooft solution$^{12}$ of 
the chiral symmetry breaking problem in gauge theories (the famous $U(1)$ 
problem). We will see that this concept will play an important role
in our proposal of the QG model whose only degrees of freedom are
the global, topological degrees of freedom. The model we propose in this 
essay shares many qualitative properties with the modern theory of strings$^6$.
\par
One may ask how we would expect to recover local physics, and in particular
classical General Relativity (GR), from a model whose only degrees
of freedom are the global, topological characteristics of the four-manifolds?
We do not have the definite answer to this question yet, but it is clear that
the four-metric on a topological four-manifold becomes a dynamical degree of
freedom at an energy scale $M_{P}$, which is of the order of the Planck mass.
Therefore, if our model displays the property of dynamical scale symmetry
breaking like Quantum Chromodynamics (QCD), then the dimensional transmutation 
phenomenon operative in QCD should take place. The dimensionless unrenormalized
coupling constant $g_{0}$ which parametrizes our model gets replaced by the 
renormalization group invariant mass parameter $M_{P}(g,\mu)$, which will 
play the role of the Planck mass in the low energy effective 
theory of gravity (GR).
\par
Below the Planck energy, the model of QG is described by an effective
low energy theory. This low energy effective theory, which is generally
covariant and causal in the classical limit, must be described by 
a lagrangian at most quadratic in the first derivatives of the metric.
There is only one such lagrangian, the Einstein-Hilbert lagrangian of General
Relativity with a possible cosmological constant. The basic property of some 
of the QG models we will discuss in this essay is that the metric becomes 
dynamical as a result of an instanton induced phenomenon.
The mechanism which is operative in the models we discuss here is in the spirit
of the induced gravity program initiated over twenty years ago by 
Zeldovich$^{2}$ and Sakharov$^{3}$, and later pursued by Adler 
and his colleagues$^{9,11}$.
\par
Quantum fluctuations of geometry and topology become important at the
Planck scale, $L_{P}=(Gh/c^3)^{1/2}=10^{-33}cm$. The argument showing 
the plausibility of Wheeler's conjecture(assertion) is based on 
the Einstein-Hilbert action or any theory of gravity with the fundamental 
mass scale, either explicit in the action or dynamically induced. It can be
easily seen, using the Feynman path integral approach and dimensional arguments,
that the probability for topology changing on distances larger than $L_{P}$
is enormously suppressed, being of order $e^{-(L/L_{P})^2}$, where $L$ is 
the scale on which topology of the four-manifold changes. Therefore, 
in the classical dynamics of the gravitational field, we do not expect 
dynamical change of topology of spacetime; topology of the $4$-manifold
{\sl is not the dynamical degree of freedom} of classical general relativity 
(GR). Before going to the more specific discussion of the topological aspects 
of Quantum Gravity (QG) let us first discuss in general terms those properties
of GR which are common with string theory.
\par
Classical general relativity is the theory of dynamics of three-geometries,
and the (closed) string theory is the theory of dynamics of one-dimensional
geometries- strings. In the hamiltonian formulation of GR (string theory) we
specify the topological three-(one-)manifold $\Sigma$ and a collection
of fields on it. For GR this is the three-metric $q_{ab}$ and its conjugate
momentum $\pi^{ab}$; for the string theory in the Polyakov formulation it is 
instead the embedding scalar field $X^{\mu}$ and the conjugate 
momentum $\Pi_{\mu}$ (for simplicity we will assume $\Sigma$ to be compact 
without boundary). The canonical variables are specified modulo the large 
gauge symmetry group, $Diff(\Sigma)$. As usual with hamiltonian systems with 
large local symmetry groups, the symmetry group $Diff(\Sigma)$ is
generated by constraints ${\sl H}_{a}$, which form (classically) a closed
algebra isomorphic to the algebra of $Diff(\Sigma)$. The classical dynamics 
is defined by the hamiltonian ${\sl H}_{0}$, which vanishes in 
reparametrization invariant (covariant) theories like GR or string theory.
${\sl H}_{0}$ evolves ${\Sigma}_{i}={\Sigma}$ to ${\Sigma}_{f}={\Sigma}$,
and the ``history'' of the geometry $\Sigma$ is the manifold 
${\sl M}={\Sigma}\times {R}$, on which the particular form of constraint algebra
induces the Lorentz structure. In the case of three-geometry, we recover in
this way the generally covariant Einstein-Hilbert action for GR.
Of course, historically this was understood the other way around.
\par
Consider the ``history'' ${\sl M}$, and a function $t$ on it.
Smooth and regular function $t$, which can be called ``local time'' defines a
one-parameter family of slices ${\Sigma}(t)$ in ${\sl M}$. In fact, $t$ is
a Morse function on ${\sl M}$ and its critical points carry information
about the topology of ${\sl M}$. If ${\sl M}$ is a product manifold
${\Sigma}\times R$, then $t$ has no critical points (i. e. points) where $dt=0$,
and starting with an arbitrary, smooth riemannian metric, using $grad(t)$ 
one can construct causal Lorentz$^{13}$ structure on ${\sl M}$.
When $t$ has critical points on ${\sl M}$ as happens for the ``trousers'' 
topology, with ${\partial {\sl M}}=\Sigma_i\cup\Sigma_f$, where 
$\Sigma_i=\Sigma_1$, and $\Sigma_f=\Sigma_2\cup\Sigma_3$, then the 
hamiltonian dynamics is not well defined because the hamiltonian ${\sl H}_0$
``does not know'' how to evolve the geometry $\Sigma_1$ beyond the splitting 
region; it vanishes identically at the critical points of the ``time'' function.
\par
It is a classical result due to Geroch$^{10}$ that nontrivial cobordisms does
not admit a smooth causal Lorentz structure. Closed timelike, or null curves
must occur on manifolds with the topology of ``trousers''. Similar phenomenon 
occurs on two-dimensional cobordisms (Sorkin,private communication).
Classical evolution of matter or gravitational fields is not well defined on 
nontrivial cobordisms, and similarly, quantum field theory(QFT) in such
backgrounds is not well defined$^{8}$. However, this does not mean that
nontrivial cobordisms are not important in QG. Nontrivial topologies may
play a fundamental role in QG, where we have to recognize the importance
of quantum fluctuations. To put it strongly, we shall propose in this essay
that all topologies of manifolds in $D=2,3,4$-dimensional QG must be 
considered$^{7}$. This might be necessary in order to have a unitary
and, presumably well defined, QG at high energies.
\par
Let us recall here the definition of an (un-)oriented cobordism. Two manifolds
are called cobordant if their disjoint union (modulo extra structure like
orientation, spin structure, complex structure etc.) bounds a closed smooth
manifold called the cobordism i. e. $\partial {\sl M}=\Sigma\cup\Sigma'$.
The last relation is an equivalence relation between closed manifolds
which transforms the space of closed manifolds into an abelian cobordism 
group $\Omega_d$, with disjoint union as the group operation. The trivial
elements of this group are all manifolds which are boundaries.
For our purpose it is important to know that $\Omega_1=\Omega_2=\Omega_3=0$.
This means that, at least in the path integral approach to QG, there are no
topological obstructions in defining a quantum mechanical amplitude for 
the $d$-geometry in the ``state'' with the topology $\Sigma_i$ to evolve to
the ``state'' with topology $\Sigma_f$. There always exist ``histories''
(cobordisms) interpolating between initial and final ``states'' $\Sigma_{i,f}$
in $D=2,3,4$-dimensional QG.
\par
To put the basic idea to work, let us now consider the instructive case 
of $D=2$ QG (or string theory). Any $1$-dimensional compact manifold $\Sigma$
is simply a disjoint union of circles $S^1$. There always exists a cobordism 
${\sl M}$ (string world-sheet) whose boundary is $\Sigma$. However, 
this cobordism is not unique unless we require that ${\sl M}$ be simply 
connected (one can always have ``holes'' in ${\sl M}$). In fact, requiring 
${\sl M}$ to be simply connected, we realize that the simplest cobordism 
is a sphere $S^2$ with a number of discs $D^2$ removed; otherwise ${\sl M}$ 
is an arbitrary Riemann surface with a number of discs removed.
However, there always exist elementary cobordisms from which we can construct 
an arbitrary cobordism by ``gluing'' together elementary cobordisms
(we can also construct in this way all compact manifolds without boundary, 
for $D=2$). We find that the elementary cobordisms in $D=2$ QG are the spheres
with two and three discs removed. In closed string field theory these are
called the propagator and vertex, respectively. Interaction in string field
theory is introduced by allowing for nontrivial cobordisms (string vertex).
\par
The quantum mechanical amplitude $A$ for quantum geometry is defined as 
a mapping between the cobordism 
${\sl M}: \partial {\sl M}=\Sigma_i\cup\Sigma_f$ with prescribed boundary 
conditions $\Phi_{|\partial {\sl M}}$ on physical fields $\Phi$ on the 
boundary $\Sigma_i\cup\Sigma_f$ and complex numbers. The boundary
conditions select a particular state from the ``asymptotic'' (Fock) Hilbert 
space for each disjoint component of $\Sigma_{i,f}$. 
$$A[{\sl M},\Sigma_{i,f},\Phi_{|\partial {\sl M}}]=
\int {\sl D}\Phi e^{-I[\Phi]} ,\eqno(1)$$
where $\Phi$ is the collection of fields depending on the model.
In string theory $\Phi=(g,X)$, where $g_{ab}$ is the world-sheet metric 
and $X^{\mu}$ is the scalar embedding field. In $D=4$ QG, the collection of 
fields $\Phi$ should include the $4$-metric $g_{\alpha\beta}$ as well
as other (matter) fields.
\par
String theory is the simplest, but complicated enough, example of a theory
in which the interaction is introduced by the {\sl topological principle}.
Witten$^6$ has constructed a quite ingeneous string field theory for open
strings. The cornerstone of his construction is the BRST formulation of the
string theory on a given cobordism (string diagram). Witten starts his
construction with the gauge-fixed string action i. e., with an action which
is not reparametrization invariant. However, this new action includes
Grassmann fields (Faddev-Popov ghosts) and has a fermionic 
symmetry---BRST invariance. When studying given theory in the BRST
formulation, one has to bear in mind that {\sl BRST invariance is 
a substitute for reparametrization invariance}. BRST invariance implies that 
there is a conserved current whose charge $Q$ is nilpotent i. e., $Q^2=0$.
The crucial property of gauge-fixed string action is that the
``energy-momentum'' tensor $T_{\alpha\beta}$ is the BRST anticommutator
$$T_{\alpha\beta}=[Q,b_{\alpha\beta}]_{+} ,\eqno(2)$$
where $b_{\alpha\beta}$ is the anti-ghost field.
This implies that the expectation value of $T_{\alpha\beta}$
vanishes on physical states:
$$\langle T_{\alpha\beta}\rangle=0 . \eqno(3)$$
This is the most important equation, because it implies that, even if
the metric $g_{\alpha\beta}$ enters explicitly the action, the quantum 
mechanical amplitudes do not depend on the metric of a given 
(string)cobordism. Also, the commutator of $Q$ and the ghost number $U$ is: 
$[U,Q]=Q$. A state is called BRST invariant if $Q$ anihilates it: $Q\psi=0$. 
The most interesting solutions of the last equation are those that cannot be 
written in the form $Q\lambda$. The equivalence classes of solutions of 
$Q\psi=0$ with a given ghost number form the BRST cohomology groups. 
Physical states of the BRST quantized system have definite ghost number $U_0$,
which depends on the particular theory. We conclude that physical
states form the BRST cohomology group $H^{U_0}$. Witten's theory$^6$ is cubic
in the string field operators. This fundamental property of Witten's theory
corresponds to the basic observation derived from cobordism theory;
there exist only two elementary cobordisms in string dynamics:
the two-holed sphere corresponding to the bare propagator which defines the
kinetic part of the second-quantized action for string fields, and the
three-holed sphere (string vertex) corresponding to the basic cubic
interaction term in the action.
\par
Encouraged by the Witten's path breaking work, some time ago$^7$ we proposed
a model of QG similar to Witten's string field theory. However, this was only 
the suggestion that one should seriously consider the possibility of 
interacting three-geometries with different topologies as the fundamental 
principle for construction of a Quantum Theory of Gravity. Our model of QG is 
one in which the interaction is introduced by the topological principle. 
In analogy with string theory we conjecture that $D=3$ QG is cubic in the
wave functional of the three-geometry. The elementary $D=3$ cobordisms
are three-manifolds with two or three boundaries, which are arbitrary Riemann 
surfaces. The fundamental object whose dynamics we will study in $D=4$ QG is
the three-geometry $\Sigma$. And here is where all the problems seem to
start. Unlike string theory, where the basic object is the one-manifold,
$S^1$, in $QG_4$ there are a plethora of possibilities for the topology of
a three-manifold. There exists an infinite number of different topological
three-manifolds, which can presumably allow for a complete classification.
What makes QG very difficult to study on the formal mathematical level is
this incredible richness of basic objects. Another problem is our relatively
poor understanding of the local and global properties of the gauge group of QG,
the diffeomorphism group $Diff(\Sigma)$.
\par
However, despite the existence of all these possible topologies $\Sigma$
in QG, it is possible to construct a model whose quantum dynamics is
independent of the metric. It is sufficient to choose the matter plus ghost
system in such a way that the energy-momentum tensor is a BRST anticommutator.
Then we study quantum mechanical amplitudes given by the the path integral
(1), where $\Phi$ is a collection of matter fields, ghosts, and the metric.
If the fundamental (gauge-fixed) action is invariant under a BRST-type
symmetry with the energy momentum tensor in the form of a BRST anticommutator,
then the amplitudes $A[\Sigma_i,\Sigma_f, \Phi|_{\partial {\sl M}}]$
will be independent of the particular metric chosen on the cobordism. 
This means that the path integral over metrics factorizes, yielding an 
irrelevant infinite constant. Assume that somehow the generator of fermionic
symmetry does not annihilate a certain state. Then the expectation value
of the energy-momentum tensor will be non-zero and and it will depend
explicitly on the metric chosen on the cobordism. In principle, we can
integrate the equation defining the variation of the effective action with
respect to the metric, $$\delta \Gamma= {1\over 2}\int d^4x
\sqrt g\delta g^{\alpha\beta}T_{\alpha\beta},\eqno(4)$$
to obtain the induced action for gravity $\Gamma[g]$.
The question we have to ask is: does there exist a mechanism which can
be responsible for the spontaneous or dynamical symmetry breaking of the
fermionic symmetry $Q$ ? This depends on the model. Witten has
recently$^5$ constructed a model with fermionic symmetry $Q$ such that the
quantum mechanical amplitudes defined by cobordisms depend only on the
topological invariants introduced by Donaldson in his studies of the theory
of four-manifolds. Witten's model is probably not very physical, but it
gives a simple physical interpretation for Donaldson's topological invariants.
>From our point of view, Witten's model has the attractive property that the $Q$
symmetry is broken by gravitational instantons (closed 4-manifolds
without boundary) i.e. by vacuum fluctuations.
\par
Let us describe the argument briefly, without going into the details of
Witten's model. The argument is based on Witten's index $Tr(-1)^F$,
familiar from supersymmetric theories. The standard argument due to Witten
shows that, if this index vanishes, then fermionic symmetry (supersymmetry)
must be spontaneously or dynamically broken. The operator $(-1)^F$ 
anticommutes with fermionic charges, $[(-1)^F,Q]_{+}=0$.
Does there exist an operator in Witten's model$^5$ which has these properties?
The answer is yes. This is the operator $e^{i\pi U}$, where $U$ is the ghost
number operator, which is defined modulo $8$ for an $SU(2)$ gauge theory
with fermionic symmetry. One can show that the generalized Witten index
for this model$^7$ is zero if the so-called formal dimension
$$d({\sl M})= 8p_1(E)-{3\over 2}(\sigma({\sl M})+\chi({\sl M})) ,\eqno(5)$$
of the moduli space of selfdual $SU(2)$ gauge connections is nonzero. $p_1(E)$
is the first Pontryagin number of the $SU(2)$ bundle $E$, and $\chi$ and
$\sigma$ are the Euler characteristic and signature of {\sl M}.
We conclude that there exist gravitational, and at the same time Yang-Mills
instantons (one can call them ``mixed'' instantons) which break the fermionic
symmetry $Q$. An example of such an instanton is the familiar $CP(2)$
instanton. ${\chi}+{\sigma}=2{B^{+}}_2$, where ${B^{+}}_2$ is the number of 
selfdual closed two-forms on {\sl M}. One can construct a number of such 
instantons. Instantons break the apparent symmetry of the classical action.
This phenomenon is quite similar to the problem of chiral symmetry
breaking in QCD. What is the physical meaning of this phenomenon?
It simply means that the theory which is generally covariant displays the
property of dynamical symmetry breaking. Witten's model is scale invariant,
because it is reparametrization invariant. Scale symmetry is dynamically
broken and the theory acquires a mass scale. At the same energy scale the
metric becomes a dynamical degree of freedom. We end up with induced
gravity as a low energy effective theory of gravity. Much more work need to 
be done before we will be able to understand properties of theories with the 
topological interaction principle. However, one thing seems to be
clear, namely that the idea of finite, nonlocal gravity beyond the Planck 
scale may shed some light on the issue of calculability of amplitudes in QG.
\par
This research was supported by NSF grants PHY 83-18350 and PHY 86-12424
with Relativity Group at Syracuse University.
\endpage
\singlespace
\centerline{REFERENCES}
\vskip .2in
\item{1)}  Wheeler, J. A., (1957), In {\sl Conference on The Role of 
Gravitation in Physics}, Chapel Hill, North Carolina, January 18-23, 1957
(WADC technical report 57-216, ASTIA document No AD 118180).
\item{2)}  Zel'dovich, Ya. B., (1967), Zh. Eksp. Teor. Fiz. Pis'ma Red.,{\bf 6}
883 [JETP Lett.{\bf 6}, 316].
\item{3)}  Sakharov, A., (1967), Dokl. Akad. Nauk SSR, {\bf 177}, 70 
[Sov. Phys. Dokl. {\bf 12}, 1040 (1968)].
\item{4)}  Hawking, S. W., (1979), In {\sl Einstein Centenary Volume},
(Cambridge University Press, Cambridge).
\item{5)}  Witten, E., (1988), {\sl Topological Quantum Field Theory},
Princeton preprint IAS-87/72; 
{\sl Topological Gravity}, Princeton preprint IAS-88/2.
\item{6)}  Witten, E., (1986), Nucl. Phys.{\bf B267}, 253.
\item{7)}  Mazur, P. O., and Nair, V. P., (February 1987), unpublished; paper
in preparation.
\item{8)}  Anderson, A., and De Witt, B., (1985), Foundations of Physics, {\bf 16}, 91.
\item{9)}  Adler, S., (1982), Rev. Mod. Phys.{\bf 54}, 729.
\item{10)} Geroch, R., (1967), J. Math. Phys.(N. Y.){\bf 8}, 782.
\item{11)} Hasslacher, B., and Mottola, E., (1980,1981),
Phys.Lett.{\bf B95}, 237; Phys. Lett.{\bf B99}, 221.
\item{12)} t'Hooft, G., (1976), Phys. Rev.{\bf D14}, 3432.
\item{13)} Sorkin, R. D. , private communication.
\end